# Superconductor Electronics Fabrication Process with MoN$_x$ Kinetic Inductors and Self-Shunted Josephson Junctions

Sergey K. Tolpygo, *Senior Member, IEEE,* Vladimir Bolkhovsky, D.E. Oates, *Fellow, IEEE,* R. Rastogi, S. Zarr, A.L. Day, T.J. Weir, Alex Wynn, and L.M. Johnson, *Senior Member, IEEE*

*Abstract*—Recent progress in superconductor electronics fabrication has enabled single-flux-quantum (SFQ) digital circuits with close to one million Josephson junctions (JJs) on 1-cm$^2$ chips. Increasing the integration scale further is challenging because of the large area of SFQ logic cells, mainly determined by the area of resistively shunted Nb/AlO$_x$-Al/Nb JJs and geometrical inductors utilizing multiple layers of Nb. To overcome these challenges, we are developing a fabrication process with self-shunted high-$J_c$ JJs and compact thin-film MoN$_x$ kinetic inductors instead of geometrical inductors.

We present fabrication details and properties of MoN$_x$ films with a wide range of $T_c$, including residual stress, electrical resistivity, critical current, and magnetic field penetration depth $\lambda_0$. As kinetic inductors, we implemented Mo$_2$N films with $T_c$ about 8 K, $\lambda_0$ about 0.51 µm, and inductance adjustable in the range from 2 to 8 pH/sq.

We also present data on fabrication and electrical characterization of Nb-based self-shunted JJs with AlO$_x$ tunnel barriers and $J_c$ = 0.6 mA/µm$^2$, and with 10-nm thick Si$_{1-x}$Nb$_x$ barriers, with $x$ from 0.03 to 0.15, fabricated on 200-mm wafers by co-sputtering. We demonstrate that the electron transport mechanism in Si$_{1-x}$Nb$_x$ barriers at $x < 0.08$ is inelastic resonant tunneling via chains of multiple localized states. At larger $x$, their Josephson characteristics are strongly dependent on $x$ and residual stress in Nb electrodes, and in general are inferior to AlO$_x$ tunnel barriers.

## I. Introduction

THERE has been a continuing progress in superconductor electronics fabrication towards increasing the number of superconducting layers and reducing the minimum size of circuit features [1]-[5]. A breakthrough into a very large scale integration (VLSI) of superconducting digital circuits has recently been made as a result. For instance, single flux quantum (SFQ) circuits containing close to one million Josephson junctions (JJs) and having circuit density over 1.3·10$^6$ JJs per cm$^2$ have been demonstrated [6],[7] using the 350-nm fabrication process SFQ5ee developed at MIT Lincoln Laboratory (MIT LL) [2]. This process utilizes resistively shunted Nb/AlO$_x$-Al/Nb Josephson junctions with critical current density $J_c$ = 0.1 mA/µm$^2$ and geometrical inductors formed on multiple superconducting wiring layers of niobium. Currently it is perhaps the most advanced fabrication process for superconductor electronics, with a theoretical maximum circuit density of about 4·10$^6$ JJs per cm$^2$ [8],[9]. However, further increase of the integration scale of SFQ circuits is challenging because of the large area of individual SFQ cells, mainly determined by the area of JJ shunt resistors and geometrical inductors [8].

There are a few fabrication technology development paths for growing the SFQ integration scale. For instance, staying with resistively shunted JJs and geometrical inductors, we can continue to decrease the minimum feature size and to increase the number of superconducting layers and layers of Josephson junctions. Transitioning to a 250-nm process for resistors, inductors, and interlayer vias would give a theoretical maximum circuit density of about 10$^7$ JJs per cm$^2$ with a single layer of junctions and proportionally higher with two layers of active devices [9]. On this route, no significant increase in the density of geometrical inductors is possible below about 250-nm linewidth and spacing because of a nearly exponential growth in the mutual inductance and cross talk between the inductors. Also, no further increase in JJ density is possible without increasing the $J_c$ of the junctions, which is required in order to reduce their area.

A more promising way is to replace externally shunted junctions with self-shunted junctions and replace geometrical inductors with kinetic inductors, and then continue to decrease the minimum feature size and increase the number of junction layers. In kinetic inductors, most of the energy is stored in kinetic energy of the supercurrent, not in the magnetic field around the inductor. Therefore, the mutual inductance between kinetic inductors remains much lower than their self-inductance even at deep submicron spacings and linewidths.

This research is based upon work supported by the Office of the Director of National Intelligence, Intelligence Advanced Research Projects Activity, via Air Force Contract FA872105C0002.

All authors are with Lincoln Laboratory, Massachusetts Institute of Technology, Lexington, MA 02421 (e-mail: sergey.tolpygo@ll.mit.edu).

Submitted to *IEEE Trans. Appl. Supercond.* on Sept. 13, 2017; revised version on Dec. 29, 2017.

The goal of this work is to develop a superconductor electronics fabrication process utilizing kinetic inductors and high-$J_c$ self-shunted Josephson junctions. We mainly concentrate here on high-$J_c$ Nb/AlO$_x$-Al/Nb junctions because so far they have shown better characteristics for SFQ digital circuits than other potential types of self-shunted JJs, e.g., Nb/Si$_{1-x}$Nb$_x$/Nb [10]-[12]. For kinetic inductors, in this paper we concentrate only on MoN$_x$ thin films, leaving other potential materials, e.g., NbN and NbTiN for another publication.

## II. KINETIC INDUCTORS FOR SFQ CIRCUITS

### A. Thin Films or Josephson Junctions?

There are two options for replacing geometrical inductors by more compact inductors with low cross talk: use the kinetic inductance of the supercurrent in a thin, $d \ll \lambda$, superconducting film $L_k = \mu_0 \lambda^2/d$; or use the kinetic inductance of the Josephson current through a junction $L_J = \Phi_0/(2\pi I_c \cos\varphi)$, a so-called Josephson inductance. Here, $\lambda$ is London magnetic-field penetration depth, $d$ the film thickness, $I_c$ the junction critical current, $\varphi$ the Josephson phase difference, and $\Phi_0$ the flux quantum. Being the same in nature, these two types of inductors utilize different structures. The former is a planar thin-film device whereas the latter is formed by a vertical stack of multiple layers. The difference practically disappears in microbridge-type thin-film JJs. So the choice of kinetic inductors should be based on technological considerations.

For a process with the minimum feature size $F$, the smallest area of a planar inductor that can be made is $F^2$. Forming an SFQ cell inductor $L_{cell}$ from a film with inductance per square $L_k$ requires $L_{cell}/L_k$ squares of the film. The area occupied by a this kinetic inductor is $A_k = (L_{cell}/L_k)F^2$ if $L_k < L_{cell}$, and $A_k = (L_k/L_{cell})F^2$ if $L_k > L_{cell}$. The density of inductors is maximized when $L_k \approx L_{cell}$. Typically, $L_{cell} \sim \Phi_0/I_c$ is about 20 pH, where $I_c \sim 100$ µA is the typical critical current of the cell junctions. The latter value is set from the cell bit error rate considerations. Therefore, the practical upper value of $L_k$ above which there is little benefit for circuit density is about 20 pH/sq and, respectively, $\lambda^2/d$ about 17 µm.

If we select a material with $\lambda$ of about 0.5 µm, we can get $(w/\lambda)^2 \sim 0.5$ using $w = F$ in the current $F = 350$ nm technology node, where $w$ is the film width. Then, the typical inductor area $A_k \approx (\Phi_0 d/\mu_0 I_c)(w/\lambda)^2$ in µm$^2$ can be less than $10d$, where $d$ is in micrometers, and getting $A_k < 1$ µm$^2$ and inductor density over $10^8$ cm$^{-2}$ would require only modestly thin films with $d$ less than about 100 nm. There are many suitable amorphous and crystalline materials, e.g., NbN, NbTiN, MoN$_x$, and such thin-film inductors can be fabricated with very high uniformity, yield, and repeatability.

The critical current of the inductor film, $I_{cf}$, has to be much larger than the cell junction $I_c \sim 100$ µA. This constrains the film's width, inductance per square $L_k$, and the ultimate circuit density. Indeed, the fundamental critical current of superconductors is the pair breaking, Ginzburg-Landau (GL) depairing critical current [13],[14] $I_{GL} = wd\Phi_0/(3\sqrt{3}\lambda^2\xi\mu_0)$; where $\xi$ is the GL coherence length, and the current distribution in the film is assumed to be uniform. Hence, the depairing critical current and the kinetic inductance are inversely proportional to each other. Their product is $I_{GL}L_k = \Phi_0 w/(3\sqrt{3}\xi)$.

In real films with $w/\xi \gg 1$, the critical current is smaller than the $I_{GL}$ and determined by the entry and motion of Abrikosov vortices from the film edges, which depend on the edge defects and current distribution. Setting $I_{cf} = \alpha I_{GL}$ with $\alpha < 1$, the $I_{cf} > I_c$ requirement results in a restriction on the film width $w > (3\sqrt{3}\xi/\alpha)L_k/20$, where $L_k$ is in pH/sq. That is, the larger $L_k$ the wider film we need to use. Assuming $\alpha \sim 0.1$ and $\xi \sim 10$ nm, this condition becomes $w/L_k > 25$ nm/pH. If we use $w = F$ to maximize the circuit density, the film kinetic inductance should be $L_k < 14$ pH/sq at $F = 350$ nm and $< 10$ pH/sq if we transition to $F = 250$ nm process node. If, to be on the safe side, we use a material with $L_k = 4$ pH/sq, the inductor area at $F = 350$ nm can be made as low as $A_k \sim 0.6$ µm$^2$.

For thin-film inductors, $L_k I_{cf}$ is adjustable by $w$. Therefore, the condition $L_k I_c > 1$ required for making quantizing kinetic inductors can be easily satisfied. However, this is not the case for junction-based inductors. Making a quantizing loop comprised of only usual Josephson junctions requires more than four serially connected JJs, $N > 4$, in order to satisfy the flux quantization condition $\sum \varphi_i = 2\pi$ before reaching the critical current of the JJs, i.e., $\varphi_i < \pi/2$, where $\varphi_i$ is the phase difference across the $i$-th junction. It would be more practical to use $N = 6$ nearly identical JJs, e.g., in a vertical stack, so that $\varphi_i \approx \pi/3$ and the loop current $I_{cir} = I_c \sin(\pi/3) \approx 0.87 I_c$ is sufficiently lower than the critical current. However, even for a 6-JJ loop, the $L_J I_c$ product (parameter $\beta_L$) is only about $2\Phi_0$, a much lower value than can be obtained by using thin-film kinetic inductors. Replacing the passive JJs with the so-called $\pi$-junctions, providing a $\pi$ phase shift across their ferromagnetic barrier, can dramatically reduce the number of required JJs [15], but creates additional technological difficulties for integrated circuit fabrication.

The Josephson critical current of the inductor JJs must be larger than the $I_c$ of the active (switching) JJs. An example of a possible fabrication process was described in [8]. There is no advantage in using a JJ stack from the point of view of the circuit density unless its area can be made smaller than the area of the thin-film inductor. From a fabrication standpoint, fabrication of thin-film inductors is much simpler and expected to have a much higher yield than fabrication of multi-junction stacks of submicron Josephson junctions. For these reasons, we concentrated in this work on the development of thin-film kinetic inductors.

TABLE I
DEPOSITION PARAMETERS AND NITROGEN CONTENT IN MoN$_x$ FILMS

| Wafer # | Power (kW) | Dep. Temp. (ºC) | N$_2$/Ar (%) | N content in Mo$_{1-x}$N$_x$ (at. %) |
|---|---|---|---|---|
| 1 | 1 | 35 | 3 | 28.6 |
| 2 | 1 | 35 | 4 | 33.3 |
| 3 | 1 | 100 | 4 | 27.3 |
| 4 | 1 | 100 | 4 | 29.0 |
| 5 | 1 | 300 | 4 | 26.3 |

## B. Fabrication of $MoN_x$ Thin-Film Kinetic Inductors

NbN and NbTiN films, typically used for kinetic inductance detectors, have magnetic field penetration depth values about 200 nm in thick films, increasing to ~ 500 nm as the thickness decreases to 3 nm; see [16]-[18] and references therein. In this work, we studied the fabrication and properties of $MoN_x$ films. Our rationale for using $MoN_x$ films was explained in [2]. It is the convenience of using the same material and deposition equipment for making both high sheet resistance films for circuit resistors and kinetic inductors by using films with different concentration of nitrogen. Previously, $MoN_x$ shunt resistors were used in superconducting circuit fabrication in [19]. In our previous work [2] we introduced $MoN_x$ kinetic inductors into our SFQ5ee process for use in SFQ circuit biasing network as large choke inductors with $L > 100$ pH required for ERSFQ biasing scheme [20]. These inductors just need to have large $L$, but the exact value and variations are not very critical. In this work, we want to implement kinetic inductors in SFQ cells where accurate targeting of the value and small parameter spread within the circuit and on wafers are very important. Therefore, we performed a careful optimization of $MoN_x$ fabrication process and characterization of $MoN_x$ inductors.

$MoN_x$ films were deposited on 200-mm wafers in a Connexion cluster tool using reactive sputtering of a Mo target in $Ar/N_2$ mixture. For process optimization, we used a fixed sputtering power, fixed Ar flow, and varied the $N_2$ flow to study a range of $N_2/Ar$ ratios from 0.01 to 0.15 in order to obtain films with different nitrogen content, critical temperature $T_c$, and resistivity. A few nanometers of titanium were used as an adhesion layer. The range of sputtering powers from 0.5 kW to 3 kW was studied. We also varied the substrate temperature from 35 °C to 400 °C. For simplicity, we targeted approximately the same film thickness, about 40 nm, for all depositions. The actual thickness was measured, and the sheet resistance data were corrected for the actual thickness.

The obtained films were studied by x-ray diffraction and secondary ion mass spectrometry (SIMS) to determine nitrogen content. Films deposited at 35 °C were found to be amorphous with a very small amount of a cubic phase with crystallite size about 2.6 nm, likely inclusions of highly disordered molybdenum and/or cubic $\gamma$-$Mo_2N$ phase. The sharpness, position, and intensity of the observed x-ray diffraction lines move closer to elemental Mo, and the crystallite size increased to about 3.7 nm and about 5.1 nm with increasing deposition temperature to 200 °C and 400 °C, respectively. The results of the SIMS measurements of nitrogen content in the deposited films are given in Table I. The composition of films with the highest $T_c \approx 8$ K obtained in this work is close to $Mo_2N$, apparently corresponding to an amorphous, or a very small grain size, version of the cubic $\gamma$-$Mo_2N$. Increasing the deposition temperature decreases nitrogen content in the films and decreases $T_c$ and electrical resistivity of the films. Therefore, room temperature depositions were mostly used to obtain more uniform amorphous films.

Electrical resistivity $\rho$ of the deposited films increases sharply with increasing $N_2$ content in the sputtering gas mix-ture above ~ 2%. It changes from values ~ 30 μΩ·cm, typical for a polycrystalline molybdenum with impurities, to values ~ 120 μΩ·cm typical for amorphous metals with very short mean free path, indicating a sharp onset of amorphization; see Fig. 1(a). The resistance ratio $R_{300}/R_{10}$ also drops in the same range of N$_2$ contents from about 3 in films with no nitrogen to about 1 above 3% N$_2$ in the flow, also indicating a transition to amorphous films.

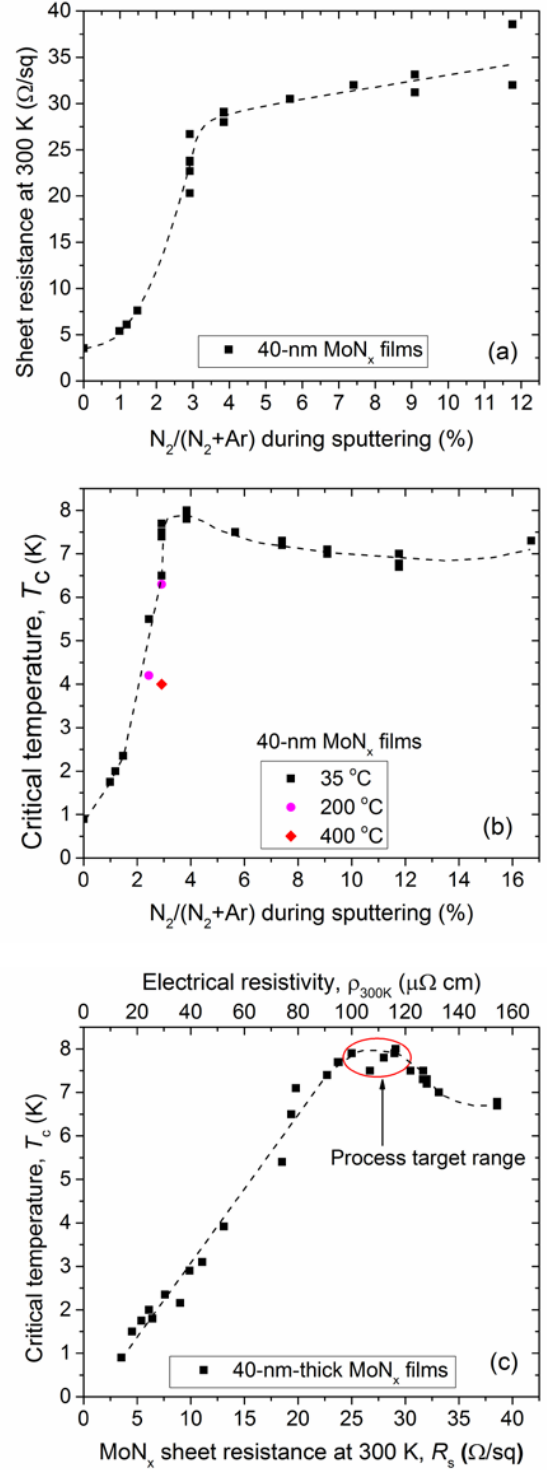

Fig. 1. (a) Sheet resistance $R_s = \rho/d$ at 300 K of deposited $MoN_x$ films with 40-nm thickness as a function of nitrogen content in $(N_2+Ar)$ sputtering gas mixture at a fixed deposition power of 1 kW and fixed deposition pressure of 3 mTorr. (b) Superconducting critical temperature $T_c$ of deposited $MoN_x$ films with 40-nm thickness as a function of nitrogen content in $(N_2+Ar)$ sputtering gas mixture. Data for a few deposition temperatures are shown. (c) Dependence of $T_c$ of deposited $MoN_x$ films on their sheet resistance. Since all films have the same thickness, this plot represents $T_c$ dependence on electrical resistivity $\rho$, shown as top axis. A plateau region near maximum $T_c$ is used as the kinetic inductance process target range. Deposition temperature was 35 °C. The dashed curves are shown only to guide the eye.

The critical temperature of MoN$_x$ films also increases sharply in the same range of N$_2$ concentrations as does the sheet resistance $R_s$; see Fig. 1(b). The $T_c$ increases to about 8 K at approximately 4% N$_2$ and then remains nearly constant, slightly above 7 K, in a very broad range of N$_2$ contents in the sputtering gas mixture. At a fixed power and N$_2$ flow, both the $T_c$ and $R_s$ (film resistivity) decrease with increasing deposition temperature, as shown by a few points in Fig. 1(b).

The critical temperature of MoN$_x$ films correlates very well with their electrical resistivity, or sheet resistance, as shown in Fig. 1(c). The $T_c$ increases nearly linearly with $\rho$ and plateaus at the highest value in the range from about 100 μΩ·cm (25 Ω/sq) to about 120 μΩ·cm (30 Ω/sq). This range was set as the process target range for kinetic inductors due to a weak sensitivity of $T_c$ to small changes in the deposition parameters.

At a fixed N$_2$ content, the film resistivity decreased with increasing the sputtering power, as shown in Fig. 2; $T_c$ decreased accordingly, as summarized in Fig. 1(c). Additional optimization of the deposition parameters was done to minimize the residual film stress. The main trends are shown in Fig. 3. At a fixed sputtering power and N$_2$/Ar flow ratio, the residual stress in MoN$_x$ films changes with deposition pressure from compressive stress at low pressures to tensile stress at high pressures. At fixed power and deposition pressure, the stress changes from tensile at low N$_2$/Ar ratios to compressive with increasing the N$_2$/Ar ratio. Optimized films had compressive stress less than 100 MPa.

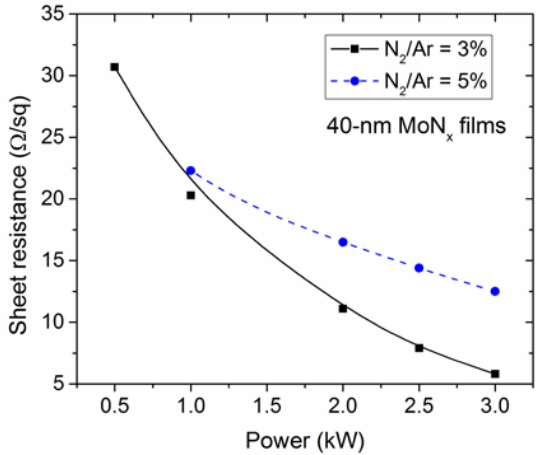

Fig. 2. The typical dependence of the sheet resistance of MoN$_x$ films on the sputtering dc magnetron power. Films of the same 40-nm thickness were deposited at a fixed pressure of 3 mTorr using various concentrations of N$_2$ in Ar shown as N$_2$/Ar flow ratio in percent.

Wafers were rotated during the deposition to improve the film thickness and sheet resistance uniformity to about 2% (one standard deviation), measured at 49 points across 200-mm wafers. Fine adjustments of film thickness were made based on measurements of kinetic inductance in unpatterned and patterned films and their critical currents.

Thermal stability of the MoN$_x$ films and their suitability for integration into a multilayer process stack with PECVD SiO$_2$ interlayer dielectric was verified by depositing 200 nm of SiO$_2$ at 150 ºC on top of the MoN$_x$ films with a 5-nm Ti adhesion layer. This sandwich was then annealed at 200 ºC for 30 and 60 minutes in N$_2$ at 2.5 Torr, and electrical characteristics were remeasured; see Fig. 4. We will refer to the films prepared this way as "annealed". The annealing temperature was set a bit higher than the maximum temperature of 190 ºC used in our SFQ fabrication processes. The annealing time was also longer than the maximum cumulative exposure time to 190 ºC temperature in our 9-metal layer process of about 40 min.

Amorphous Mo$_2$N films in the optimum range of resistivity corresponding to the $T_c$ plateau in Fig. 1(c) were stable at 200 ºC and showed very small, less than 2%, change (increase) in resistivity. We attribute this increase to an interac-tion of Mo$_2$N films with Ti adhesion layer. MoN$_x$ films with resistivities below ~ 100 μΩ·cm showed a progressive decrease of resistivity with annealing time, up to ~ 10%, which was larger for lower-$\rho$ films, possibly a result of partial recrystallization.

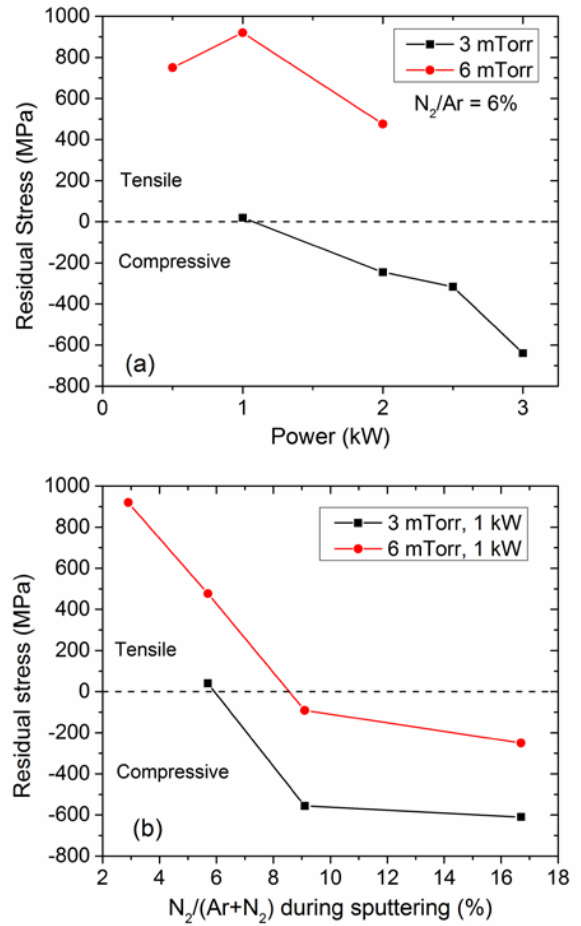

Fig. 3. (a) The typical dependences of the residual stress in deposited MoN$_x$ films on the sputtering dc magnetron power at a fixed N$_2$/Ar flow ratio for two deposition pressures. (b) Residual stress dependences on the nitrogen content in the sputtering gas mixture at a fixed power of 1 kW and pressures of 3 mTorr and 6 mTorr, respectively. The films target thickness was 40 nm.

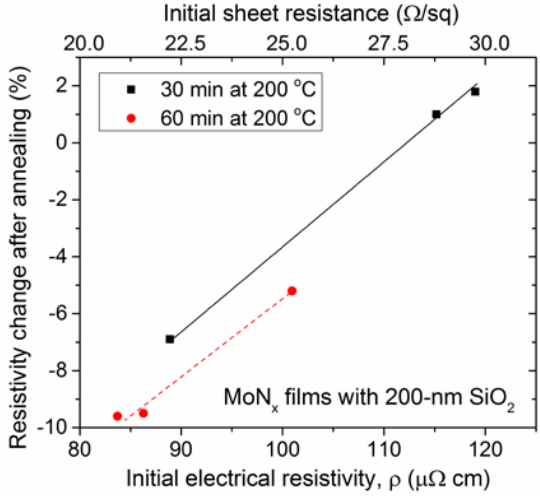

Fig. 4. Relative change (in percent of the initial value) in electrical resistivity of MoN$_x$ films deposited at 35 ºC, then coated with 200-nm layer of SiO$_2$ at 150 ºC and annealed at 200 ºC for 30 and 60 minutes. A 5-nm adhesion layer of Ti was used between MoN$_x$ and PECVD oxide.

### C. Microwave Measurements of $\lambda(T)$ in MoN$_x$ films

Magnetic field penetration depth and its temperature dependence $\lambda(T)$ of unpatterned MoN$_x$ films were measured using a dielectric-resonator technique shown schematically in resonant frequency down, and microwave losses in the film reduce the $Q$-factor. The relative change of the resonance frequency with temperature is proportional to the change in the inductance of two films $L(T) = 2\mu_0\lambda_{\text{eff}}(T)$ and, since the frequency shifts are very small, can be expressed as

$$f_r(T)/f_r(T_0) = 1 - g^{-1}\cdot[\lambda_{\text{eff}}(T) - \lambda_{\text{eff}}(T_0)], \quad (1)$$

where $g$ is a geometric factor of the resonator and $\lambda_{\text{eff}}(T)$ is the effective penetration depth. In general, $\lambda_{\text{eff}} = \lambda\coth(d/\lambda)$, which gives $\lambda_{\text{eff}} = \lambda^2/d$ for a very thin $d \ll \lambda$ and $\lambda_{\text{eff}} = \lambda$ in the opposite case. In the simplest two-fluid model, the $T$-dependence of the penetration depth is

$$\lambda(T) = \lambda_0/[1 - (T/T_c)^4]^{1/2}. \quad (2)$$

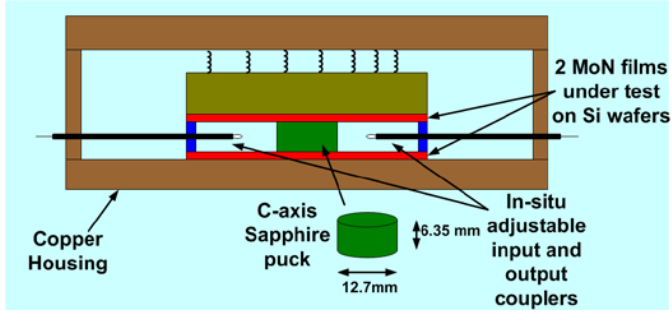

Fig. 5. Dielectric resonator test setup for measuring microwave properties of superconducting films. The resonance frequency was 10.7 GHz. The copper housing was placed in a vacuum can with He exchange gas. Temperature was varied from 4.2 K to $T_c$ of the films.

Fig. 5; see, e.g., [21]-[23]. For these measurements, 200-mm wafers with deposited MoN$_x$ films were diced into multiple octagonal-shaped wafers with 2-inch inscribed diameter. These smaller wafers were attached to the flat sides of a sapphire cylindrical resonator with 6.35-mm height and 12.7-mm diameter. The resonator quality factor and the resonant frequency $f_r$ were measured in the $T$-range from $T_0$ = 4.24 K to $T_c$. The dielectric resonator TE$_{011}$ mode was used with the resonance frequency about 10.7 GHz and quality factor $Q$ over 34,000. The inductance of the superconducting films shifts the

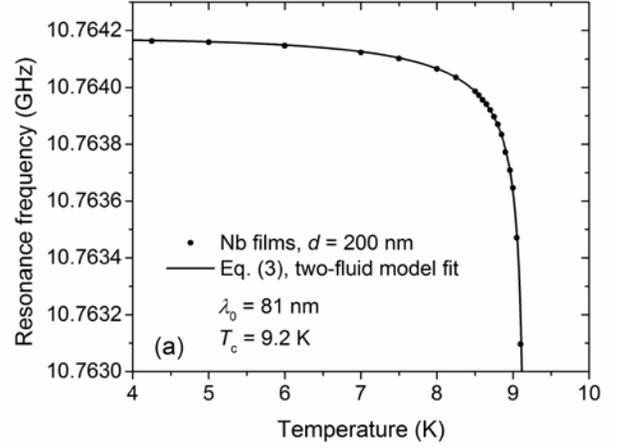

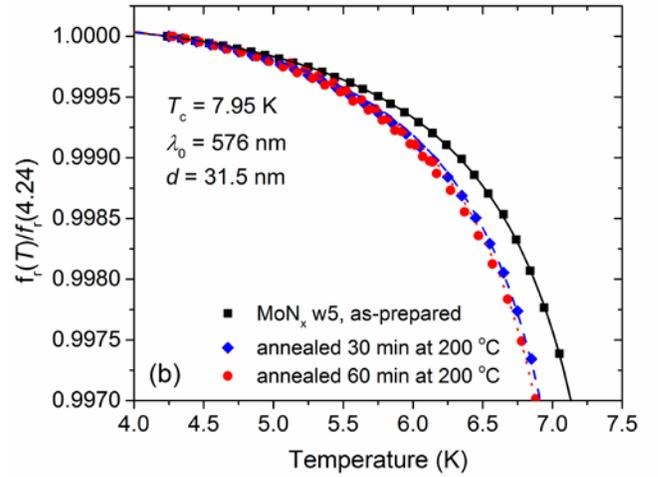

Fig. 6. (a) Dielectric resonator measurements of a 200-nm niobium film. Solid curve is the fit to (3) giving $\lambda_0$ = 81 nm. (b) Dielectric resonator measurements of the typical as-prepared MoN$_x$ film. Fit to (4) is shown by the solid black curve, giving $T_c$ = 7.95 K and $\lambda_0$ = 576 nm. The film was also measured after annealing in N$_2$ at 200 ºC for 30 min, resulting in $T_c$ = 7.68 K and $\lambda_0$ = 567 nm (blue dashed curve), and 60 minutes, resulting in $T_c$ = 7.63 K and $\lambda_0$ = 563 nm (red dotted curve).

Then, the $f_r(T)$ change with respect to the $f_r$ at the liquid He temperature, $f_r(T_0)$, is given by

$$f_r(T)/f_r(T_0) = 1 - g^{-1}\lambda_0\{[1 - (T/T_c)^4]^{-1/2} - [1 - (T_0/T_c)^4]^{-1/2}\} \quad (3)$$

in the thick film $\lambda(T) << d$ limit, and

$$f_r(T)/f_r(T_0) = 1 - g^{-1}(\lambda_0^2/d)\{[1-(T/T_c)^4]^{-1} - [1-(T_0/T_c)^4]^{-1}\} \quad (4)$$

in the thin-film $\lambda(T) >> d$ limit.

The geometrical factor of the resonator was calculated, and an additional calibration was done using deposited 200-nm Nb films. The zero-$T$ penetration depth $\lambda_0 \equiv \lambda(0)$ and $T_c$ were determined by fitting the experimental data to (3) or (4). Thus obtained $T_c$ values agreed within $\pm 0.1$ K with the values obtained from the dc resistive transitions measured on the same films.

Fig. 6 shows the typical results of the measurements on Nb films and on the typical MoN$_x$ films. The 200-nm Nb films are in the thick-film regime, except for very near to $T_c$, and (3) fits the $T$-dependence of the resonance frequency exceptionally well, as shown in Fig. 6(a). In contrast, thin MoN$_x$ films with large $\lambda_0$ are in the thin-film regime in the entire temperature range, and (4) fits the experimental data very well as shown in Fig. 6(b).

We measured films deposited at different conditions and annealed films. We have found that MoN$_x$ films with the maximum $T_c$, corresponding to films with composition Mo$_2$N and electrical resistivity around 110 $\mu\Omega\cdot$cm in Fig. 1(c), have the lowest penetration depth $\lambda_0 \approx 0.54$ $\mu$m. The $\lambda_0$ increases as $T_c$ decreases on both sides of the maximum, i.e., either with increasing electrical resistivity to higher values or decreasing the resistivity. This indicates that the concentration of superconducting electrons $n_s$ is maximized in Mo$_2$N films with the maximum $T_c$ since the London magnetic field penetration depth $\lambda_L$ scales inversely with the Cooper pair density $\lambda_L^2 = m/(\mu_0 n_s e^2)$. Apparently, $n_s$ decreases on both sides of the optimum independent of whether the resistivity of the films increases or decreases.

We noted larger variation of $\lambda_0$ in films with reduced $T_c$ than in the optimal, "stoichiometric" Mo$_2$N films, apparently indicating larger sensitivity of the Cooper pair density $n_s$ to the nitrogen content and other film fabrication details. Similarly, 60-min anneals of nonoptimal films at 200 ºC sometimes resulted in increased value of the measured $\lambda_0$, suggesting some changes in the morphology of the films because changes in $T_c$ and resistivity were minor.

The resonator quality factor $Q$ at $T = 4.24$ K decreased strongly with decreasing $T_c$ of the films, as shown in Fig. 7, likely a result of an exponential growth in concentration of normal quasiparticles with increasing $T/T_c$ ratio. The low $Q$ values were making measurements of $\lambda$ in lower-$T_c$ films less accurate than in the films with the highest $T_c$ and $Q$.

In order to compare the penetration depth data in different MoN$_x$ films, we recall that $\lambda_0$ in the microscopic theory of superconductivity is given by $\lambda_0^2 \approx \lambda_L^2(0)(\xi_0/l)$ in the dirty limit $l << \xi_0$, which is certainly applicable to our amorphous superconducting films. Here, $l$ is the mean free path and $\xi_0 \equiv \hbar v_F/\pi\Delta_0$ is the Bardeen-Cooper-Schrieffer (BCS) coherence length, and $\Delta_0$ the energy gap; see, e.g., [24]. Then, simple transformations and the BCS relationship $\Delta_0 = 1.76 k_B T_c$ give

$$\lambda_0 = (\hbar\rho/\pi\mu_0\Delta_0)^{1/2} \approx 0.105(\rho/T_c)^{1/2}, \quad (5)$$

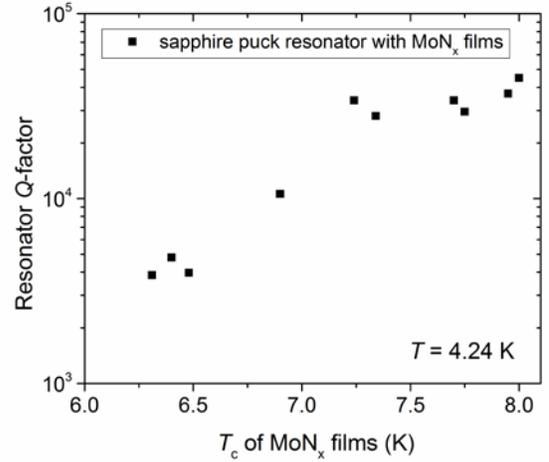

Fig. 7. Dependence of the quality factor $Q$ at $T = 4.24$ K of a sapphire resonator used for magnetic field penetration depth measurements on $T_c$ of the attached MoN$_x$ films. $Q$ diminishes nearly exponentially with lowering $T_c$ due to the exponential growth of the concentration of normal quasiparticles with increasing $T/T_c$.

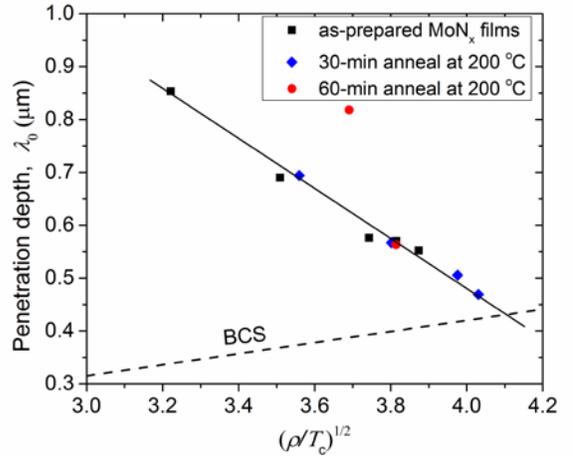

Fig. 8. Dependence of magnetic field penetration depth extrapolated to zero temperature, $\lambda_0$, in MoN$_x$ films on their electrical resistivity and $T_c$. The BCS theory dependence (5) is shown by the dashed line. Solid line is to guide the eye. Some of the films annealed at 200 ºC for 60 min showed large changes in $\lambda_0$, as shown by one of the red circle located above the solid line. The data are from the dielectric resonator measurements, see text.

where $\lambda_0$ is in micrometers, $\rho$ in $\mu\Omega\cdot$cm, and $T_c$ in Kelvin.

The measured values of $\lambda_0$ are shown in Fig. 8 as a function of $(\rho/T_c)^{1/2}$ for the as-prepared and annealed films with various $T_c$ values. Surprisingly, we observe the opposite dependence of $\lambda_0$ on $(\rho/T_c)^{1/2}$ to the BCS relationship (5) shown by the dashed line. The penetration depth $\lambda_0$ decreases with increasing $(\rho/T_c)^{1/2}$ instead of increasing. Because $T_c$ grows approximately linearly with $\rho$, $T_c = \alpha\rho$ with $\alpha \sim 0.08$ K/($\mu\Omega$ cm), as shown in Fig. 1c, we would expect $\lambda_0$ to be the same in all films $\lambda_0 \approx 105/\sqrt{\alpha}$, and independent of $(\rho/T_c)^{1/2}$ since $T_c$ dependence cancels out. This is clearly not the case. For the films with the optimum value $\rho = 110$ $\mu\Omega\cdot$cm and $T_c = 8$ K, we get $\lambda_0 = 390$ nm from (5), a value about 30% lower than the measured. This may indicate that $n_s$ is almost a factor of two lower than the normal electron number density entering into $\rho$, e.g., due to some pair breaking effects.

A reduction in the superfluid density with respect to the BCS values was also observed in polycrystalline MoN films obtained by atomic layer deposition, using terahertz frequency-domain spectroscopy [25]. Usually, increased values of the penetration depth in thin films are attributed to formation of a Josephson network of weakly coupled grains; see, e.g., [26] and references therein. In this case, the effective penetration depth and kinetic inductance are determined by the Josephson penetration depth $\lambda_J = \Phi_0/(2\pi\mu_0 I_c)$ between the grains and Josephson inductance, respectively.

Josephson junction networks are usually formed in thin films because of the columnar growth, *e.g.*, in transition-metal films and their nitrides such as Nb and NbN, and/or grain boundary oxidation, *e.g.*, in Nb and Al films, and high-$T_c$ superconductors; see [27] for a review and references. The evidence for Josephson network behavior in granular $MoN_x$ films deposited by thermal evaporation was presented in [28]. However, critical currents in $MoN_x$ films formed by reactive sputtering, see below and [29], and our SEM data partially presented in [9] do not indicate granular structure in our $MoN_x$ films, although do show it in $NbN_x$ films. However, further discussion of these issues goes beyond the scope of this paper.

### D. $L_k$ and Critical Current of Patterned $MoN_x$ Films

$Mo_2N$ films with the target thickness of 40 nm were incorporated into SFQ5ee process stack with eight Nb metal layers [2]. The films were patterned using 248-nm photolithography and etched using $SF_6/O_2$ mixture with 60/40 sccm flow at 10 mTorr in a high-density plasma etcher at 600 W rf power and 60 W bias power.

For inductance measurements, a dc-SQUID based technique described in [30] was used. Patterned $Mo_2N$ strips of various widths from 0.5 to 2 μm and various lengths formed inductors of dc SQUIDs. The film inductance was calculated from the period of SQUID modulation. We verified that unshielded films of different width and length, as well as films placed under a Nb ground plane to form inverted microstrips, gave nearly the same inductance per square $L_k$, indicating negligible contribution of the geometrical inductance of the films. The average sheet inductance obtained on 40-nm films at 4.2 K was 8.2 pH/sq, corresponding to $\lambda(4.2\ K) = 511$ nm and $\lambda_0 = 490$ nm, about 10% lower value than the one we obtained from microwave measurements on unpatterned films; see II.C.

The average $L_k$ values obtained on wafers fabricated during the two-year period since the launch of the SFQ5ee process are shown in Fig. 9. In the actual process stack, $Mo_2N$ film thickness was adjusted to get the required sheet inductance value. The observed run-to-run variations of $L_k$ are about 4% (one standard deviation) and apparently represent a cumulative effect of run-to-run variations of the $Mo_2N$ films thickness and composition.

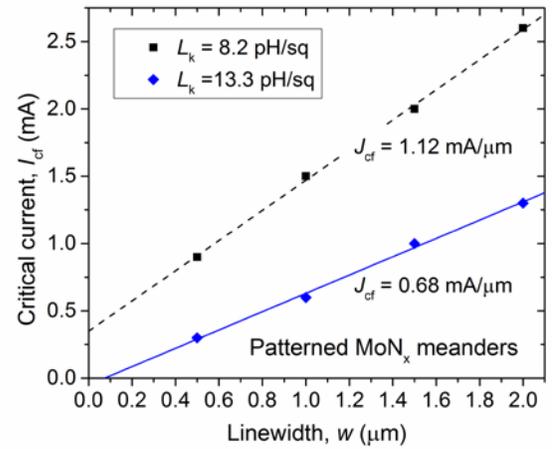

Fig. 10. The typical critical currents of patterned $MoN_x$ films as a function of width $w$. The $L_k J_{cf}$ product is about 9.1 pH mA/μm or about $4.4\Phi_0$/μm.

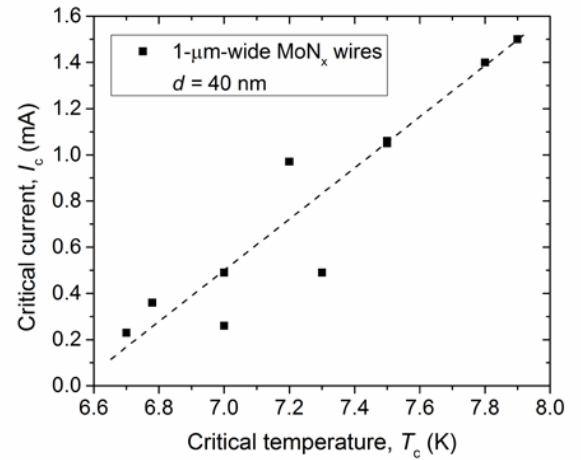

Fig. 11. Critical current of patterned 1-μm wide $MoN_x$ films with different values of $T_c$. The targeted film thickness is 40 nm. The maximum critical current density observed is $3.8 \cdot 10^6$ A cm$^{-2}$, a factor of 10 lower than the critical current of Nb films with the same thickness. This reduction is caused by a much lower concentration of Cooper pairs $n_s$ in $MoN_x$ than in Nb, which follows from a much larger penetration depth $\lambda_0$.

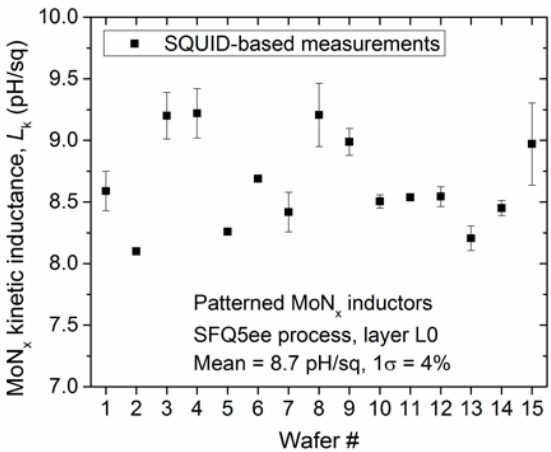

Fig. 9. Wafer-averaged kinetic inductance of patterned $Mo_2N$ films from SQUID-based measurements. The wafers were fabricated during a two-year period since the launch of the SFQ5ee process and represent the regular process runs. The targeted film thickness is 40 nm.

The critical current of patterned MoN$_x$ films, $I_{cf}$ was measured on the same chips as used for SQUID measurements of $L_k$. The films were patterned as meanders with length to width ratios $N_{sq}$ from about 100 to 500 squares. The typical $I_{cf}$ versus width dependences are shown in Fig. 10. From the linear slopes we determined the critical current density per unit width $J_{cf}$ shown in Fig. 11.

From a circuit design standpoint, the important characteristic of a kinetic inductor is the $LI_{cf}$ product, which for the inductor with length $\ell$ is $LI_{cf} = \ell L_k J_{cf}$. From II.A, we expect the $L_k J_{cf}$ product to be the film material property independent of the film thickness and width

$$L_k J_{cf} = \alpha \Phi_0 / (3\sqrt{3}\xi), \tag{6}$$

where $\alpha < 1$ characterizes reduction of the critical current with respect to the Ginzburg-Landau depairing current. We have found the product $L_k J_{cf} = 4.4\Phi_0/\mu m$ for our films and indeed independent of $L_k$. This value gives $\xi/\alpha = 44$ nm. Hence, a 1-μm square of Mo$_2$N film provides a factor of two larger product of inductance and critical current than a stack of six JJs.

The critical current of the films correlates with $T_c$, as shown in Fig. 11, and reaches a maximum in the same range of film resistivities as does the $T_c$. However, the maximum critical current density observed in Mo$_2$N films is a factor of 10 lower than in Nb films with the same thickness. This difference is caused by a much lower concentration of Cooper pairs $n_s$ in Mo$_2$N than in Nb as indicated by a much larger penetration depth $\lambda_0$.

## III. SELF-SHUNTED JOSEPHSON JUNCTIONS

### A. High-$J_c$ Nb/AlO$_x$-Al/Nb Junctions

It is well known that the most commonly used Nb/AlO$_x$-Al/Nb Josephson junctions become self-shunted at sufficiently high values of $J_c \sim 1$ mA/μm$^2$ [31]. This self-shunting is caused by a greatly enhanced subgap conductance by multiple Andreev reflections via defects in the tunnel barrier [32]. The $J_c$ required to achieve self-shunting is fabrication dependent. In [33] we investigated a range of $J_c$ from 0.1 to 1 mA/μm$^2$ for our technology. We selected a $J_c \approx 0.6$ mA/μm$^2$ as the process target. This $J_c$ is high enough to provide sufficient self-shunting with McCumber-Stewart parameter $\beta_c$ from about 2 to 3, but still gives JJs with sufficiently low parameter spreads and good uniformity on 200-mm wafers to be suitable for VLSI circuits. The typical current-voltage (I-V) characteristics are shown in Fig. 12; see also [33]. More detailed results will be published elsewhere.

### B. Nb/Si:Nb/Nb Josephson Junctions

Another group of self-shunted JJs are junctions of an *SNS* or *SSmS* type using a normal-metal or semiconductor barrier. In order to get $I_c R_N$ product values high enough to be suitable for high-speed SFQ circuits, highly disordered metals or doped semiconductors are used as a barrier, usually with compositions close to the metal-insulator transition [11],[12]. Some promise was shown by Nb-based JJs with amorphous Si (a-Si) barriers doped by Nb, W, or other dopants forming deep impurity levels close to the middle of Si band gap. They were proposed and investigated a long time ago; see, e.g., [10],[34], and references therein to even earlier work. Recently, JJs with a-Si:Nb barriers [35],[36] found applications in programmable Josephson voltage standards and waveform synthesizers developed at NIST USA and PTB Germany. These applications require very large-area junctions with critical currents over 1 mA and low $I_c R_N$ products $\sim 0.1$ mV, JJ counts in only a few hundred thousand range, and do not require fabrication on large-size wafers.

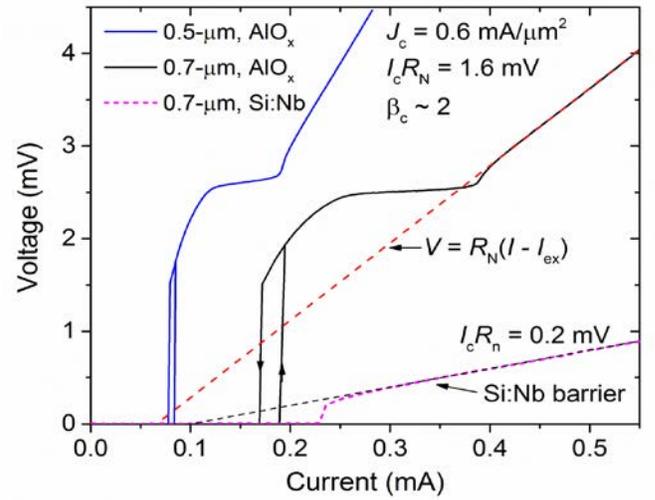

Fig. 12. Current-voltage characteristics, from left to right, of 0.5 and 0.7-μm diameter Nb/AlO$_x$-Al/Nb junctions, and 0.7-μm-diameter Nb/Si$_{1-x}$Nb$_x$/Nb junctions with deposited 10-nm barrier of Si doped with $x=0.09$. The AlO$_x$-based JJs are nearly self-shunted with a small hysteresis corresponding to $\beta_c$ range from $\sim 2$ to 3. The JJs with Si$_{1-x}$Nb$_x$ barrier at $x = 0.09$ are overdamped and nonhysteretic. Note a very large difference in the characteristic voltage $V_c = I_c R_N$ of these two types of JJs, making JJs with doped Si barriers less attractive for applications in high-speed SFQ circuits.

The requirements for superconducting digital circuits are quite different: submicron-size junctions with $I_c \sim 0.1$ mA and lower, $I_c R_N$ products over 1 mV, low JJ parameters spread, JJ counts over a million, and good uniformity over large-size wafers. Therefore, we investigated fabrication of JJs with doped-Si barrier on 200-mm wafers in order to evaluate their uniformity and suitability of their parameters to VLSI SFQ circuits.

Nb/Si$_{1-x}$Nb$_x$/Nb trilayers were deposited on 200-mm oxidized Si wafers in-situ in a Canon-Anelva deposition system. Nb-doped Si barriers were deposited using co-sputtering of Si and Nb targets with 100-mm diameters. Wafers were rotating during the deposition; the obtained thickness uniformity was better than 1%. The obtained uniformity of Nb and Si$_{1-x}$Nb$_x$ sheet resistance $\rho/d$ was in the range from 0.4% to 0.9% depending on Nb film stress, and 1.7% and 1.4% for Si$_{1-x}$Nb$_x$ films with $x = 0.09$ and 0.1, respectively. Full optimization of JJs' deposition requires optimization of the barrier thickness and Nb content because the critical current of SSmS and SNS junctions exponentially depends on the barrier thickness $d$ and coherence length of the barrier material $\xi_N$, which in turn depends on the doping level and temperature. Instead, we set the barrier thickness at 10 nm, because it was estimated as the lowest thickness that could provide better than 2% run-to-run reproducibility of the barrier thickness on 200-mm wafers. The amount of Nb doping $x$ in Si was varied from 3 to 15 atomic percent, based on the results of the prior work [11],[12],[34]-[37].

After deep-UV photolithography, junctions were etched using $Cl_2$-based chemistry and planarized using CMP of a PECVD $SiO_2$ interlayer dielectric as described in [1]. Circular junctions with design diameters from 350 nm to a few micrometers were fabricated and studied. The junction resistance uniformity was characterized at room temperature using a wafer prober, measuring resistance of 440 JJs with diameters 0.5, 0.7, 1, and 1.4 μm per 5-mm chip, and from 9 to 45 chips per wafer. A much smaller subset of JJs was characterized at 4.2 K by $I$-$V$ measurements.

Nb/$Si_{1-x}Nb_x$/Nb junctions with Nb concentration in the barrier of 8 at. % and below do not show any signs of Josephson current at 10 nm barrier thickness and 4.24 K. Instead, current-voltage $I(V)$ characteristics of the junctions are highly nonlinear, as shown in Fig. 13.

The observed $I(V)$ characteristics are indicative of electron transport by resonant tunneling via chains of localized states, so-called Lifshitz' resonance-percolation trajectories [41]-[44]. For instance, at $x = 0.05$ and 4.24 K, the $I(V)$s are very well described by a polynomial form

$$I = G_0V + G_2V^{7/3} \qquad (7)$$

in the range of voltages $|V| < 40$ mV studied; see Fig. 13. The second term in (7), which dominates at $V > 10$ mV, is due to inelastic tunneling via chains consisting of two localized states as had been predicted a long time ago [44] and since then observed experimentally in numerous types of junctions with disordered insulators, including metal-insulator-metal structures with a-Si [45]. Fits to (7) give the specific tunneling conductance $G_0A^{-1}$ of about 7 mS/μm$^2$, close to the value in Nb/AlO$_x$-Al/Nb tunnel junctions with $J_c \sim 10$ μA/μm$^2$ and $\sim 1$-nm thick barriers, where $A$ is the junction area.

Junctions with $x = 0.03$ display a much stronger nonlinearity than (7). Their $I(V)$ dependences at 4.24 K can be described well by

$$I = G_0V + G_2V^{7/3} + G_3V^{7/2}, \qquad (8)$$

where the third term, which becomes dominant at $V > 25$ mV, represents a contribution of inelastic resonant tunneling via chains with three localized states [44]. Due to a larger average distance between Nb impurities, the first term in (8) becomes almost negligible at $x = 0.03$, $G_0A^{-1} \sim 7$ μS/μm$^2$. The two-impurity chains contribution $G_2A^{-1}$ also decreases from about 3 A V$^{-7/3}$ μm$^{-2}$ at $x = 0.05$ to about 0.05 A V$^{-7/3}$ μm$^{-2}$ at $x = 0.03$. That is, with decreasing doping, electron transport through a-Si barrier favors chains with larger number of resonant impurities. More details on the current transport through a-Si:Nb barriers at low doping will be given elsewhere.

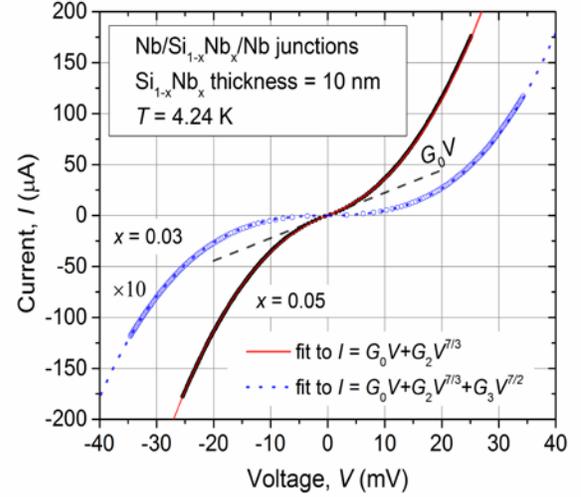

Fig. 13. The typical current-voltage characteristics of 0.7-μm-diameter junctions with 10-nm thick $Si_{1-x}Nb_x$ barriers at low levels of Nb doping: $x = 0.03$ (blue open circles) and 0.05 (black solid points). Note the scale difference: current was multiplied by 10x to show the $I(V)$ for $x=0.03$ in this scale. Solid red curve is the best fit to (7), which is indistinguishable from the data points. The dashed line shows the cumulative contribution of direct tunneling and elastic resonant tunneling, $G_0V$ term in (7). Dotted blue curve is the best fit to (8) for the junction with $x=0.03$; see text. Note a much stronger $I(V)$ nonlinearity in the junction with lower doping level.

$Si_{1-x}Nb_x$ barriers with Nb concentration of 9 at. % resulted in $J_c \sim 0.6$ mA/μm$^2$. Higher doping levels, $x > 0.1$, resulted in $J_c$s too high and not suitable for SFQ circuits. Using the sheet resistance of deposited $Si_{1-x}Nb_x$ films, we determined in-plane resistivity $\rho = R_sd$ to be 4.27 mΩ cm and 3.14 mΩ cm at $x = 0.09$ and 0.1, respectively. Hence, the specific barrier resistance (in the direction perpendicular to the junction plane) $R_{sq} = R_sd^2$ was expected to be about 0.43 and 0.31 Ω μm$^2$, respectively, i.e., a factor of 10 lower than for AlO$_x$-based tunnel barriers at $J_c = 0.6$ mA/μm$^2$. The fabricated junctions showed a factor of 2x higher specific resistances of 0.78 and 0.7 Ω μm$^2$ for $x = 0.09$ and 0.1, respectively. The on-chip and on-wafer resistance uniformities were found to be nearly the same and equal to 1.2% and 1.5% for 0.7-μm and 1-μm JJs, respectively.

Despite relatively small changes in junction resistance with doping at these levels of $x$, the Josephson critical current density $J_c$ was found to strongly depend on the composition $x$ and the residual stress in Nb electrodes. The $J_c$ changes from 0.62±0.05 mA/μm$^2$ to 1.42±0.07 mA/μm$^2$ with $x$ increasing from 0.09 to 0.10.

We have found that residual stress in Nb films has a profound effect on $J_c$ of Nb/$Si_{1-x}Nb_x$/Nb junctions. For instance, a change in residual stress of Nb films from about −100 MPa to −700 MPa resulted in the $J_c$ reduction from 1.12 to 0.62 mA/μm$^2$ at $x = 0.09$ and from 2.07 to 1.42 mA/μm$^2$ at $x = 0.10$, a 60% reduction on average. A similar change in the stress results in only about 10% change in the $J_c$ of Nb-based junctions with AlO$_x$ tunnel barriers. These results indicate that accurate $J_c$ targeting and reproducibility needed for a stable VLSI process with Nb/$Si_{1-x}Nb_x$/Nb junctions require a very precise process control over both Nb doping within a 0.1% or better level and over Nb film deposition parameters.

The typical *I-V* characteristics of a 0.7-μm-diameter JJ with 9% doping is shown in Fig. 12 for a comparison with $AlO_x$-based tunnel junctions with about the same $J_c$. The characteristic voltage $V_c = I_c R_N$ is only 0.2 mV, which is a factor of 8 less than in $Nb/AlO_x$-Al/Nb JJs. This indicates that the maximum clock frequency of SFQ circuits utilizing JJs with $Si_{1-x}Nb_x$ barrier will be much lower than in circuits with $Nb/AlO_x$-Al/Nb JJs.

By inspecting the published *I-V* characteristics of junctions with doped Si barriers, we noted that $I_c R_N$ products reported for nonhysteretic JJs with various barrier thicknesses, doping types and levels in [10]-[12],[34]-[37] and other publications do not exceed 0.36 mV, the same value as the $I_c R_N$ of critically damped $Nb/AlO_x$-Al/Nb junctions at $J_c = 10$ μA/μm$^2$. Therefore, it appears to be not possible to optimize junctions with doped Si barriers by varying $d$ and $x$ to produce nonhysteretic junctions with values of $J_c$ in the practical range, $J_c \leq 2$ mA/μm$^2$, and $I_c R_N$ products comparable to values obtained in self-shunted tunnel junctions, about $\pi\Delta_{Nb}/2 \approx 2$ mV.

Indeed, getting close to the latter value requires barriers with $d < \xi_N$. At $x = 0.09$, $\xi_N \approx 5.6$ nm according to our estimate. Making barriers thinner than that would increase $J_c$ to unpractically high values requiring nanoscale junctions. Their properties would be similar to nano-constrictions considered in [38]-[40], which by the way do not require any barrier. Thinning the barrier would also dramatically compromise JJ uniformity and reproducibility. Reducing the doping level would decrease $\xi_N$ to the resonant tunneling length in Si, about 1 – 2 nm, driving JJs towards the tunneling regime realized in $AlO_x$ barriers, and making them hysteretic. Moreover, nm-thick deposited Si barriers are expected to be less uniform and reproducible than $AlO_x$ grown by self-limiting oxidation.

Note also the presence of a large excess current, $I_{ex}$, in *I-V* characteristics, as indicated by dashed lines in Fig. 12. The excess current in SSmS junctions also follows from a current transport mechanism by resonant tunneling via chains of localized impurity centers [42],[46]. Fluctuations in the number of these resonant centers and transparency of these trajectories due to compositional fluctuations are responsible for sample-to-sample variations of the junction conductance and Josephson critical current.

The $I_c$ variation of 0.7-μm and 1-μm junctions with deposited $Si_{1-x}Nb_x$ barriers on-chip and on-wafer have been found so far to be a factor of about 3x larger than for $Nb/AlO_x$-Al/Nb JJs with the same sizes and the same average $J_c$ of 0.6 mA/μm$^2$. Similarly, the run-to-run repeatability of $I_c$ was inferior to $AlO_x$-based junctions. A more detailed description of the results on $Nb/Si_{1-x}Nb_x/Nb$ junction fabrication and characterization will be published elsewhere, whereas we also plan to do more experiments with submicron JJs with doped Si barriers. For the process integration with kinetic inductors, we concentrated on using $Nb/AlO_x$-Al/Nb due to their superior characteristics.

To conclude, we note that there are other types of junctions which exhibit self-shunting, have sufficiently high $J_c$, and relatively high $I_c R_N$ values, e.g., SINIS [47], SNIS [48],[49], etc., see a recent review [50]. However, discussion of their pros and cons goes beyond the scope of this manuscript because we are not considering them for the use in our fabrication processes.

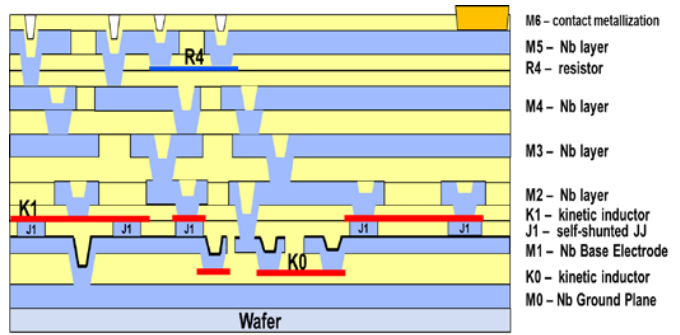

Fig. 14. The target process cross section, showing two possible positions of the $Mo_2N$ kinetic inductors, labeled K0 and K1, in the process stack. The simplified version has six planarized niobium layers and one layer of resistors. The full process stack is expected to have eight planarized niobium layers and an additional layer of bias kinetic inductors near the top of the stack.

## IV. PROCESS INTEGRATION

### A. *$Mo_2N$ Kinetic Inductor Layer Placement*

In the SFQ5ee process, kinetic inductors are used for circuit biasing and, as a result, are placed as the first layer in the process stack and interconnected by the first layer of Nb [2]. In our new process we want to use kinetic inductors as SFQ cell inductors. This requires placing them as close to JJs as possible in order to minimize parasitic inductance of interlayer vias. Therefore, we are completely changing the process stack and layer sequence, as shown in Fig. 14, while preserving planarization and other features of the SFQ5ee process.

The first superconducting layer in the process is a Nb ground plane. It is followed by a $Nb/AlO_x$-Al/Nb trilayer. For placing and interconnecting kinetic inductors, we are investigating two options: a) right below the trilayer base electrode (BE) and using BE interconnections, shown as layer K0; b) above JJs with interconnections by the next Nb layer, shown as layer K1. In the SFQ5ee process, the latter place is occupied by the shunt resistors, but it is available in this newer process node due to the use of self-shunted JJs. Depending on the circuit design requirements, we could implement two layers of kinetic inductors K0 and K1. Alternatively, we can use a layer of resistors instead of K0 to facilitate additional JJ shunting if required. For circuit biasing, we implement the regular layer of resistors, R4 in Fig. 13, which is identical to the SFQ5ee process, but placed differently in the stack. Finally, if bias kinetic inductors are required, they will be placed above the shown stack and two more Nb layers will be added to interconnect them and provide a sky plane (top ground plane) for circuits. All layers in our process are fully planarized. This allowed us successfully fabricate two fully independent layers of Josephson junctions, similar to the double-JJ-layer process demonstrated in [51]. Availability of multiple junction layers should enable increasing the integration scale of superconductor electronic circuits in a 3D manner.

*B. Kinetic Inductance Target*

Since many components of SFQ circuits may use relatively low inductances, e.g., about 2 pH in Josephson transmission lines (JTLs), implementing 8-pH-per-square kinetic inductors would not make much sense. Therefore, we increased the target thickness of our $Mo_2N$ films to 120 nm in order to decrease $L_k$ below 3 pH/sq and, accordingly, increase $J_{cf}$ by a factor of 3x to about 3 mA/μm. This should allow us to use 0.35-μm-wide inductors and have enough margin in their critical current.

## V. Conclusion

We presented a detailed investigation of the superconducting properties of thin $MoN_x$ films, such as critical temperature, critical current, magnetic-field penetration depth and its *T*-dependence, as well as a detailed optimization of the fabrication parameters in order to implement amorphous $Mo_2N$ kinetic inductors in superconducting integrated circuits. We have found that the product of kinetic inductance and critical current of a 1-$μm^2$ $Mo_2N$ film is about $4.4\Phi_0$, which is a factor of 28 larger than the $LI_c$ product for Josephson junctions, $\Phi_0/2\pi$. This makes thin $Mo_2N$ films more attractive for integration into VLSI circuits as kinetic inductors than stacked Josephson junctions requiring a much more complex fabrication process.

We investigated $Nb/Si_{1-x}Nb_x/Nb$ junctions with 10-nm thick barriers deposited by co-sputtering and *x* in the range from 0.03 to 0.15. We showed that electron transport through $Si_{1-x}Nb_x$ barriers at low doping levels *x* and low temperatures is due to inelastic resonant tunneling via chains of localized states with two and three localized states. At *x* above about 0.08, $Nb/Si_{1-x}Nb_x/Nb$ junctions become nonhysteretic Josephson junctions with relatively low $I_cR_N$ products, a few times lower than in $Nb/AlO_x$-Al/Nb Josephson junctions at the same critical current density $J_c$. Strong dependence of $J_c$ on Nb doping level *x* and residual stress in Nb electrodes was also observed.

We compared $Nb/Si_{1-x}Nb_x/Nb$ junctions to self-shunted $Nb/AlO_x$-Al/Nb Josephson junctions at the $J_c$ value used in our current SFQ5vhs process node [33], about 0.6 mA/$μm^2$. Based on a limited set of Josephson junctions with deposited amorphous $Si_{1-x}Nb_x$ barriers studied thus far (about a dozen 200-mm wafers) with 10-nm thickness and $x = 0.09$ and 0.1, we concluded that $AlO_x$-based tunnel junctions have superior parameters, at the same $J_c$, for applications in high-speed SFQ digital circuits.

We developed an integration scheme and a process stack to implement $Mo_2N$ kinetic inductors in SFQ logic/memory cells along with up to eight planarized Nb layers.

## Acknowledgment

We are very grateful to Corey Stull for his help in developing deposition of $MoN_x$ films, and to Mr. Hideki Fujii and many colleagues at Canon-Anelva for depositing Nb-doped Si films and $Nb/Si_{1-x}Nb_x/Nb$ trilayers used in this work. We would also like to thank Eric Dauler, Marc Manheimer, and Scott D. Holmes for their interest in and support of this work.

The views and conclusions contained in this publication are those of the authors and should not be interpreted as necessarily representing the official policies or endorsements, either expressed or implied, of the ODNI, IARPA, or the U.S. Government. The U.S. Government is authorized to reproduce and distribute reprints for Governmental purposes notwithstanding any copyright annotation thereon.